\documentclass[shortAfour,sageh,times]{sagej}
\usepackage{moreverb,url,nameref,float}
\usepackage[colorlinks,bookmarksopen,bookmarksnumbered,citecolor=blue,urlcolor=blue]{hyperref}
\newcommand\BibTeX{{\rmfamily B\kern-.05em \textsc{i\kern-.025em b}\kern-.08em
T\kern-.1667em\lower.7ex\hbox{E}\kern-.125emX}}
\def\volumeyear{\the\year{}}
\def\journalname{Quantum Economics and Finance}
\usepackage{amsmath,amssymb,mathtools}

\begin{document}

\runninghead{Guarcello, Dubkov, Valenti, and Spagnolo}

\title{Noise-Assisted Metastability: From L\'evy Flights to Memristors, Quantum Escape, and Josephson-based Axion Searches}

\author{Claudio Guarcello\affilnum{1,2}, Alexander A. Dubkov\affilnum{3}, Davide Valenti\affilnum{4}, Bernardo Spagnolo\affilnum{4,5}}

\affiliation{
\affilnum{1}Dipartimento di Fisica ``E.\ R.\ Caianiello'', Universit\`a degli Studi di Salerno, I-84084 Fisciano, Salerno, Italy
\affilnum{2}INFN, Sezione di Napoli, Gruppo Collegato di Salerno -- Complesso Universitario di Monte S.\ Angelo, I-80126 Napoli, Italy
\affilnum{3}Radiophysics Department, Lobachevsky State University, 603950 Nizhniy Novgorod, Russia
\affilnum{4}Dipartimento di Fisica e Chimica ``E.\ Segr\`e'', Group of Interdisciplinary Theoretical Physics, Universit\`a degli Studi di Palermo, I-90128 Palermo, Italy
\affilnum{5}Stochastic Multistable Systems Laboratory, Lobachevsky University, 603950 Nizhniy Novgorod, Russia
}

\corrauth{Bernardo Spagnolo, Dipartimento di Fisica e Chimica “E. Segr\`e”, Group of Interdisciplinary Theoretical Physics, Università degli Studi di Palermo, I-90128 Palermo, Italy.}

\email{bernardo.spagnolo@unipa.it}

\begin{abstract}
Many-body and complex systems, both classical and quantum, often exhibit slow, nonlinear relaxation toward stationary states due to the presence of metastable configurations and environmental fluctuations. Nonlinear relaxation in a wide variety of natural systems proceeds through metastable states, which arise in condensed-matter physics as well as in fields ranging from cosmology and biology to high-energy physics. Moreover, noise-induced phenomena play a central role in shaping the dynamics of such systems far from equilibrium.
   This review develops a unifying perspective centered on noise-assisted stabilization and the statistical properties of metastable dynamics. We first discuss escape processes driven by L\'evy flights in smooth metastable potentials, emphasizing the emergence of nonmonotonic residence-time behavior. We then connect these concepts to stochastic resistive switching in memristive devices, where noise-induced effects can enhance stability and reproducibility. We further examine driven dissipative quantum bistability, showing how the interplay between external driving and system–environment coupling reshapes escape pathways and lifetimes. Finally, we outline how switching-time statistics in current-biased Josephson junctions can provide an experimentally accessible strategy for axion detection, based on an axion-induced resonant-activation signature.

\end{abstract}


\maketitle

\section{Introduction}

In this paper, we briefly review recent results that collectively elucidate the nontrivial role of noise, dissipation, and external driving in stabilizing and controlling metastable states in both quantum and classical systems. We emphasize the constructive role of noise in the dynamics of complex systems and highlight a unifying theme: the counterintuitive enhancement of stability induced by fluctuations, which challenges the traditional view of noise as purely detrimental.

Metastable states are a characteristic feature of nonlinear complex systems and govern relaxation, switching, and escape phenomena in physics, chemistry, and materials science.
The effects of Gaussian noise on metastability are well understood, most notably through Kramers' theory and its generalizations to nonlinear potentials and nonequilibrium initial conditions~\citep{Agudov2003}. However, many physical systems are instead subject to non-Gaussian fluctuations characterized by rare but large events. Such fluctuations are naturally described by L\'evy noise, which exhibits heavy-tailed statistics and formally infinite variance. The work reported by \cite{Dubkov2025} addresses the fundamental question of how L\'evy noise affects the stability of metastable states. Remarkably, L\'evy noise enhances metastable stability and gives rise to asymptotic behavior in the zero-noise limit that is fundamentally different from the Gaussian case. This finding generalizes the concept of noise-enhanced stability to non-Gaussian stochastic dynamics. In particular, exact results have been obtained for the mean residence time (MRT) of a particle moving in an arbitrary smooth potential with a sink under L\'evy noise with arbitrary L\'evy index $\alpha$ and noise intensity. Moreover, a closed expression in quadrature for the MRT has been derived for L\'evy flights with $\alpha = 1$ (Cauchy noise) in a cubic metastable potential, analytically demonstrating the enhancement of metastable stability induced by L\'evy noise~\citep{Dubkov2025}. These results broaden the theoretical framework and underscore the universality of noise-assisted stabilization phenomena.

The constructive role of noise is not only of theoretical relevance but also emerges clearly in solid-state devices. Memristors, for example, are multistable systems whose switching dynamics is intrinsically stochastic, as consistently observed in experiments. To employ memristors as functional elements in resistive random access memory (RRAM) and neuromorphic architectures, it is essential to deepen our understanding of the resistive state switching process, accounting for multistability, internal and external noise sources, and metastable states in the transient nonlinear dynamics of these nonequilibrium systems. In \cite{Agudov2020}, \cite{Filatov2022} and \cite{Koryazhkina2022}, the first experimental evidence of noise-enhanced stability in ZrO$_2$(Y) memristors was reported. A stochastic model introduced by \cite{Agudov2020} predicts a nonmonotonic dependence of the relaxation time on fluctuation intensity, in good agreement with experimental observations~\citep{Koryazhkina2022}. These works demonstrate that noise can either slow down or accelerate switching depending on its nature and strength, and that noise can play a genuinely constructive role in memristive systems. Furthermore, another paradigmatic noise-induced phenomenon---stochastic resonance---has been experimentally observed in metal-oxide memristive devices~\citep{Mikhaylov2021}.

Metastability and noise-assisted phenomena are equally central in quantum systems. Metastable states appear in a wide range of quantum platforms, from superconducting circuits and cold atoms to chemical reactions and condensed-matter systems. Traditionally, dissipation and noise have been regarded as mechanisms that accelerate decay and destroy quantum coherence. However, growing evidence indicates that dissipation and quantum fluctuations, when properly engineered, can instead be exploited as a resource to control quantum dynamics. \cite{Valenti2018} explores this paradigm shift by analyzing how external driving and dissipation can stabilize quantum metastable states rather than acting solely as sources of decoherence. The study demonstrates that the interplay between coherent driving, dissipation, and quantum fluctuations can significantly prolong the lifetime of metastable states, revealing a genuine quantum analogue of the noise-enhanced stability observed in classical systems~\citep{Agudov2003}. Specifically, it is shown that increasing the system--environment coupling drives a transition in the escape dynamics: from a regime where the escape time is strongly controlled by the external driving and exhibits resonant peaks and dips, to a regime where the escape time becomes frequency independent, characterized by a single peak followed by a steep decrease. The escape time displays a nonmonotonic dependence on bath coupling, temperature, and driving frequency, providing clear evidence of quantum noise-enhanced stability and quantum resonant activation in the investigated system.

Finally, these concepts also open new perspectives in fundamental physics and sensing technologies. The search for axions and axion-like particles as viable dark matter candidates remains one of the most compelling challenges in modern physics. Conventional detection strategies---such as those exploiting the Primakoff effect in microwave cavities---rely on the conversion of axions into photons in strong magnetic fields, but suffer from limitations in sensitivity and broadband coverage over wide axion mass ranges. In this context, superconducting devices, and in particular Josephson junctions (JJs), have emerged as promising alternatives owing to their intrinsic sensitivity to weak perturbations and their tunable dynamical properties~\citep{Grimaudo2022}. The phenomenon of resonant activation in a current-biased Josephson junction was proposed by \cite{Grimaudo2022} as a mechanism for axion detection. The key conceptual advance lies in coupling the axion field to the JJ dynamics such that the effective escape dynamics of the phase particle in the tilted washboard potential are modified by the presence of the axion field. This coupling manifests as a nonmonotonic dependence of the mean switching time---from the superconducting to the resistive state---on the ratio between the axion energy and the Josephson plasma energy. When this ratio approaches unity, a resonant condition enhances the switching probability, providing a distinct axion-induced signature in the switching-time statistics that could be exploited experimentally.

\section{L\'evy-noise-induced enhancement of stabilization of metastable states}
\addcontentsline{toc}{section}{L\'evy-noise-induced stabilization}

In this section, we discuss escape from metastability under symmetric $\alpha$-stable L\'evy noise, emphasizing the noise-enhanced stability scenario that emerges when analyzing the mean residence time in a prescribed interval. We summarize the role of boundary conditions and geometric parameters defining the observation window. The resulting parameter dependence provides a compact characterization of L\'evy-induced stabilization effects and of their signatures in trajectory data.\\

Escape from a metastable state is a standard framework for noise-driven transitions in physics, chemistry, and complex systems.
In the classical setting of Gaussian diffusion, the mean escape time typically decreases monotonically with increasing noise intensity, in line with Kramers-type physics~\citep{Kra40}.
Over the years, however, a number of nonequilibrium scenarios have shown that the lifetime of a metastable state can be enhanced by noise over an intermediate range of intensities, producing a pronounced maximum of a suitable mean time scale.
This counterintuitive behavior is usually discussed under the denomination of noise-enhanced stability (NES)~\citep{Agu01,Dub04,Fia05,Hur06}.

A natural question is how these ideas change when the driving fluctuations are non-Gaussian and dominated by rare large events.
A canonical model is provided by symmetric $\alpha$-stable L\'evy flights, whose increments have heavy tails and allow arbitrarily long jumps.
Such processes appear in many contexts and admit complementary descriptions, namely a Langevin equation driven by L\'evy white noise, and a fractional Fokker--Planck equation with a nonlocal space derivative~\citep{Lev37,Gne54,Met00,Dub08}.
In barrier-crossing problems, L\'evy flights can modify the notion of activation because a single large jump may overcome the barrier region, and stationary measures may deviate from the usual Gibbs form~\citep{Che05,Che07,Dyb07,Imk09,Imk10}.
In barrier-crossing problems, L\'evy flights can overcome the barrier through rare long jumps, so that escape is not solely governed by diffusive exploration up to the barrier top, as in the Gaussian case~\citep{Che05,Che07,Dyb07,Imk09,Imk10}

L\'evy flights and, more generally, $\alpha$-stable fluctuations provide a standard modeling route for dynamics dominated by rare large increments~\citep{Dub08,Zab15}.
Within condensed-matter and material settings, L\'evy-type statistics have been discussed in charge and phase-transport problems, including graphene-related platforms, as a possible mechanism behind impulsive, non-thermal switching events~\citep{Bri14,Gat16,Gua17,Kis19,Fon23,Fon24}.
Signatures consistent with heavy-tailed fluctuations have also been reported in optical and semiconductor contexts, from photoluminescence in doped samples and interstellar scintillation to transport and optical response in nanocrystal quantum dots~\citep{Lur10,Sem12,Sub14,Bol03,Bol05,Bol06,Gwi07,Nov05,Kun00,Kun01,Shi01,Mes01,Bro03}.
Beyond these examples, $\alpha$-stable noise models are routinely used in applied settings where impulsive disturbances and outliers are relevant, including quasiballistic heat conduction, telecommunications and network interference, and vibration-based condition monitoring~\citep{Ver15I,Ver15II,Moh15,Upa16,Yan03,Bha06,Cor10,Tsi95,Sub15,Sho15,Kar20,LiYu10,Cho14,Saa15,Ely16,Whi84,McF84}. In this direction, Josephson-based threshold detection schemes have been proposed to access non-Gaussian, $\alpha$-stable, current fluctuations by leveraging the sensitivity of switching/escape observables to rare, large excursions~\citep{Gua19,Gua21PRR,Gua21CSF,Gua24}.
More broadly, heavy-tailed jump statistics have also been discussed in other application areas, including geophysical and climate variability, biological motion and search, and financial time series; in those cases the main commonality is the presence of intermittency and rare, high-impact events, rather than a shared microscopic mechanism.

More broadly, L\'evy-type statistics have been discussed and used as effective models in biological motion and search~\citep{WestDeering94,Brockmann06,Sims08,Reynolds09} and in financial time series~\citep{Mantegna95}. 
Related heavy-tailed descriptions have also been considered in geophysical and atmospheric variability~\citep{Shlesinger95,Ditlevsen99}, where the common theme is intermittency and rare, high-impact events rather than a shared microscopic mechanism.

In this section, we discuss the framework to quantify metastable trapping under symmetric L\'evy noise using residence-time observables~\citep{Dubkov2025}.
Rather than focusing exclusively on first-passage events, we consider the mean residence time (MRT) within a given interval, which provides a robust measure of how long trajectories spend inside a prescribed observation window.
A key advantage is that the MRT can be derived from the fractional transport equation without imposing high-barrier or weak-noise approximations.
We then specialize to the analytically tractable case of Cauchy noise ($\alpha=1$) in a cubic metastable potential, where an explicit quadrature formula for the MRT can be obtained and used to demonstrate a finite, nonmonotonic NES-like behavior.

\subsection{Model}

We consider an overdamped particle in a smooth potential $V(x)$ under symmetric $\alpha$-stable L\'evy noise,
\begin{equation}
\frac{dx}{dt} = -V'(x) + L_\alpha(t), \qquad 0<\alpha<2,
\label{eq:langevin_review}
\end{equation}
where $L_\alpha(t)$ is a white L\'evy noise source~\citep{Dub05,Dub08}.
The transition probability density $P(x,t|x_0,0)$ obeys the fractional Fokker--Planck equation
\begin{equation}
\frac{\partial P}{\partial t}
= \frac{\partial}{\partial x}\!\left[V'(x)\,P\right]
+ D_\alpha\,\frac{\partial^\alpha P}{\partial |x|^\alpha},
\label{eq:ffpe_review}
\end{equation}
where $D_\alpha$ sets the noise intensity, with typical spread $(D_\alpha t)^{1/\alpha}$, and
$\partial^\alpha/\partial|x|^\alpha$ is the Riesz fractional derivative~\citep{Met00,Jes99,Wes97,Yan00}.

\begin{figure*}[ht]
\centering
\includegraphics[width=\linewidth]{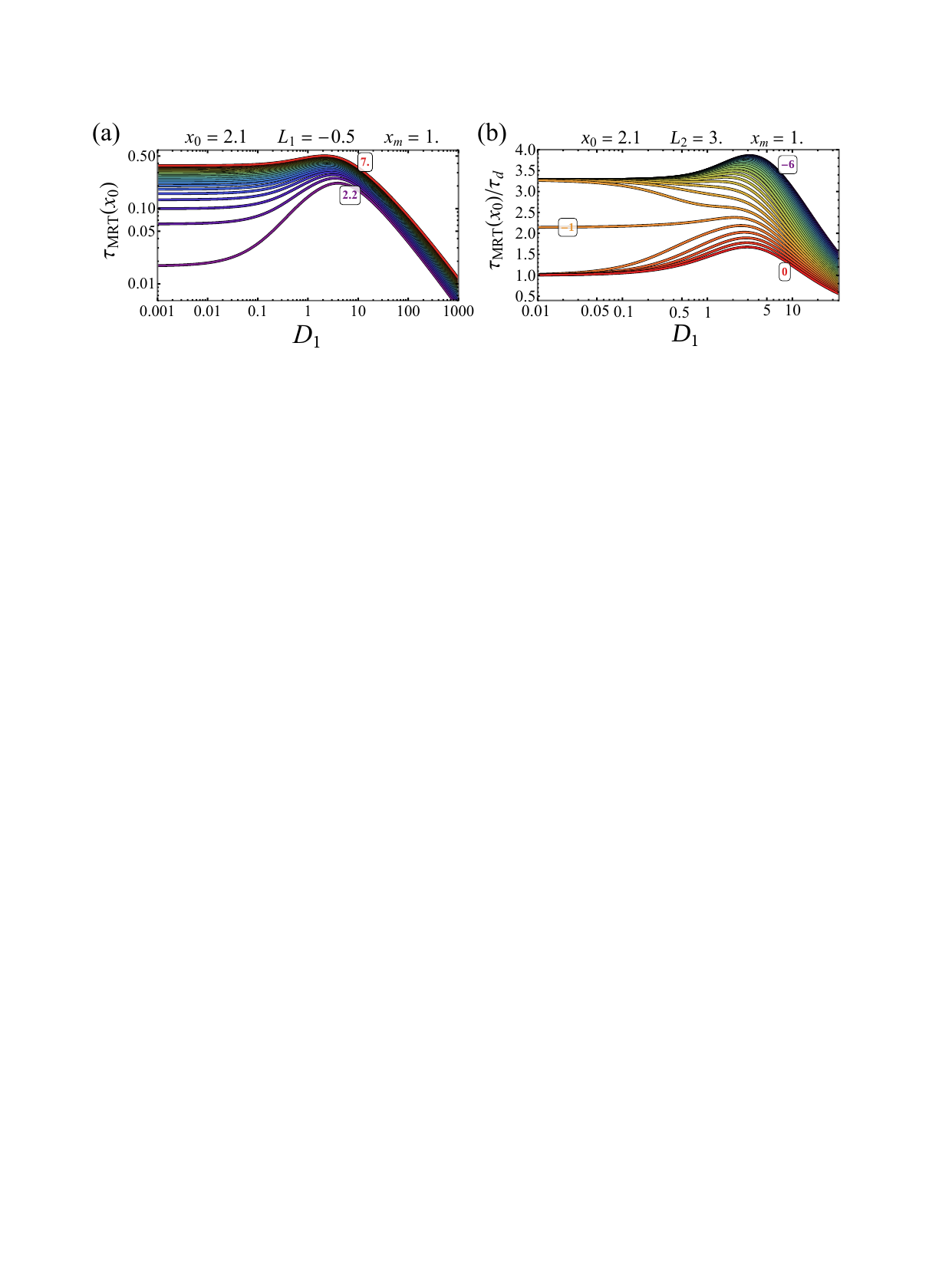}
\caption{MRT for Cauchy noise ($\alpha=1$) in the cubic metastable potential $V(x)=-x^3/3+m^2x$ with $m=x_m=1$ and initial condition $x_0=2.1$.
(a) MRT $\tau_{\mathrm{MRT}}(x_0)$ versus noise intensity $D_1$ for $L_1=-0.5$ and $L_2\in[2.2,7]$ in steps of $0.2$ (log-log scale, $D_1\in[10^{-3},10^{3}]$), highlighting the NES maximum and the large-$D_1$ power-law decay.
(b) Normalized MRT $\tau_{\mathrm{MRT}}(x_0)/\tau_d(x_0)$ versus $D_1$ for $L_2=3$ and $L_1\in[-6,0]$ in steps of $0.2$, showing the NES maximum for all explored $L_1$ and the characteristic ``duck-bill'' structure associated with boundary-controlled trapping under L\'evy flights.
Here $\tau_d(x_0)$ denotes the deterministic transit time to $L_2$ for the noiseless overdamped dynamics, when this time is finite.}
\label{fig:two_panel_MRT}
\end{figure*}

To quantify metastable trapping, we introduce an observation window $(L_1,L_2)$ and define the residence time~\citep{Dubkov2020}
\begin{equation}
T(x_0)=\int_0^\infty 1_{(L_1,L_2)}\!\big(x(t)\big)\,dt,
\label{eq:RT_review}
\end{equation}
with
\begin{equation}
1_{(L_{1},L_{2})}(y) =
\begin{cases}
1, & y \in [L_{1},L_{2}],\\
0, & \text{otherwise}.
\end{cases}
\end{equation}
The corresponding MRT is

\begin{equation}
\tau_{\mathrm{MRT}}(x_0)=\langle T(x_0)\rangle
=\int_0^\infty\!dt\int_{L_1}^{L_2}\!P(x,t|x_0,0)\,dx.
\label{eq:MRTdef_review}
\end{equation}

Operationally, $\tau_{\mathrm{MRT}}$ measures how strongly the metastable region traps stochastic trajectories.
For a purely deterministic overdamped dynamics (no noise), the MRT either reduces to a deterministic transit time, if the trajectory crosses the interval, or becomes ill-defined when deterministic barrier crossing is impossible.
In contrast, under L\'evy noise, rare long jumps can produce barrier crossing even at very small noise intensities, which makes the small-noise behavior of $\tau_{\mathrm{MRT}}$ quite different from the Gaussian case.

A successful  strategy is to introduce the time-integrated propagator
$Z(x,x_0)=\int_0^\infty P(x,t|x_0,0)\,dt$.
Integrating Eq.~\eqref{eq:ffpe_review} over time yields an exact equation for $Z$~\citep{Dubkov2025},

\begin{equation}
\frac{d}{dx}\!\left[V'(x)\,Z\right] + D_\alpha \frac{d^\alpha Z}{d|x|^\alpha} = -\delta(x-x_0),
\label{eq:Zeq_review}
\end{equation}

which can be analyzed efficiently in Fourier space.
This approach leads to closed expressions for $\tau_{\mathrm{MRT}}(x_0)$ in terms of auxiliary functions defined by the Fourier-transformed problem (see \cite{Dubkov2025} for further details).

\subsection{Results}

The final closed expression for $\tau_{\mathrm{MRT}(x_0)}$ and the differential equation for the auxiliary function $G(k,z)$, defined as the derivative with respect to $x_0$ of the Fuorier transform of the function $Z(x,x_0)$, are given by

\begin{equation} 
\tau_{\mathrm{MRT}}(x_0)\!=\!\!\int_{x_0}^{\infty}\!\! \text{Re} \!
\left\{\int_{0}^{\infty}\!\! G(k,z)\frac{e^{-ikL_2}\! - \!e^{-ikL_1}}{\pi i k} dk\right\}\! dz,
\label{eq:MRTgeneral_review}
\end{equation}

and 

\begin{equation}
V'\left(-i\frac{d}{dk}\right) G - iD_{\alpha} k^{\alpha-1} G = exp^{ikx_0}.
\label{eq:Geq_review}
\end{equation}

These equations \eqref{eq:MRTgeneral_review} and \eqref{eq:Geq_review} are the exact relations useful to calculate the MRT of a symmetric L\'evy flights with arbitrary index $\alpha$ and noise intensity parameter $D_{\alpha}$ in a smooth potential profile with a sink at $x = \infty$.

A generic metastable potential can be modeled by the cubic form

\begin{equation}
V(x)=-\frac{x^3}{3}+m^2 x,
\label{eq:cubic_review}
\end{equation}

which exhibits a local minimum at $x=-m$ and a barrier top (local maximum) at $x\equiv x_m=+m$ with $m>0$.
We focus on the Cauchy case $\alpha=1$, which is analytically tractable while still capturing the hallmark feature of L\'evy flights, namely heavy-tailed jumps.
In this case, the Fourier-space problem reduces to a linear ordinary differential equation and yields an explicit expression for the MRT in quadratures

\begin{align}
\tau_{\mathrm{MRT}}(x_0) \!&=\!
\frac{D_{1}}{\pi}\!\!\int_{x_0}^{\infty}\!\!\left[
A\!\left(\frac{z\!+\!\lambda_2}{\lambda_1}\right)\!\!+\!B\right]\!
\frac{dz}{(z^2\!-\!m^2)^2\!+\!D_1^2}\nonumber\\
&\quad+\frac{D_{1}}{\pi}\int_{x_0}^{\infty}\!
\ln\!\left|\frac{z-L_1}{z-L_2}\right|\frac{dz}{(z^2-m^2)^2+D_1^2}
\nonumber\\
&\quad
+\int_{x_0}^{L_2}\frac{(z^2-m^2)\,dz}{(z^2-m^2)^2+D_1^2},
\label{eq:mrt_quadrature}
\end{align}

with

\begin{eqnarray}
A&=&\arctan\frac{\lambda_2+L_2}{\lambda_1}-\arctan\frac{\lambda_2+L_1}{\lambda_1},\\
B&=&\frac{1}{2}\ln\frac{\lambda_1^2+(L_2+\lambda_2)^2}{\lambda_1^2+(L_1+\lambda_2)^2},
\label{eq:AB_def}
\end{eqnarray}
and $\lambda=\lambda_1+i\lambda_2$ depending on $(m,D_1)$ through the complex root structure of the Cauchy-driven problem \citep{Dubkov2025}. The equations \eqref{eq:MRTgeneral_review}, \eqref{eq:Geq_review} and \eqref{eq:mrt_quadrature} are the main results of this study.

Equation~\eqref{eq:mrt_quadrature} provides an operational link between a measurable time scale, $\tau_{\mathrm{MRT}}(x_0)$, and the control parameters, $D_1$ and $(L_1,L_2)$.
The systematic feature is a finite nonmonotonic dependence of $\tau_{\mathrm{MRT}}$ on $D_1$, with a pronounced maximum over an intermediate range of noise intensities, see Fig.~\ref{fig:two_panel_MRT}.
This maximum represents the NES signature for heavy-tailed fluctuations.
A compact physical picture is based on competing contributions: increasing $D_1$ promotes repeated entry and re-entry into the metastable region, which increases the accumulated residence time, while larger $D_1$ also increases the incidence of long jumps that eject trajectories rapidly from the sink, which reduces residence.

A practical point is that the NES nonmonotonicity of the mean residence time persists over broad parameter ranges and remains well defined within the chosen observation window. In particular, rare long jumps provide an additional pathway for leaving (and re-entering) the region, so the small-noise behavior and the very meaning of “residence” depend explicitly on how the interval boundaries are set. This dependence becomes particularly transparent when $\tau_{\mathrm{MRT}}$ is scanned versus $D_1$ while varying either $L_2$ at fixed $L_1$, or $L_1$ at fixed $L_2$.

The two-panel representation in Fig.~\ref{fig:two_panel_MRT} is a convenient summary of boundary-controlled trapping.
Panel (a) shows $\tau_{\mathrm{MRT}}(x_0)$ versus $D_1$ for a family of right boundaries $L_2$ at fixed $L_1$.
In this plot, varying $L_2$ changes the effective distance to absorption and therefore the weight of long-jump exits, while keeping the left boundary fixed.
The persistence of an NES maximum across the explored $L_2$ range indicates that the enhancement is not tied to a single geometric choice, and the log-log scale makes the large-$D_1$ power-law decay directly visible.

Panel (b) focuses on the normalized observable $\tau_{\mathrm{MRT}}(x_0)/\tau_d(x_0)$ at fixed $L_2$ while scanning the left boundary $L_1$.
Here, $\tau_d(x_0)$ denotes the deterministic transit time to the right boundary in the noiseless overdamped dynamics. It is worth noting that both the maxima and the overall magnitude of the curves increase as the left boundary $L_1$ is shifted to smaller values, i.e., as the basin of attraction of the metastable state is enlarged, leading to an increase in the normalized MRTMRT \citep{Dubkov2025}. The resulting family of curves displays the characteristic “duck-bill” profile previously reported by \cite{Dubkov2025}, with the NES maximum systematically shifting as $L_1$ is varied.
This behavior can be interpreted as a boundary-controlled crossover between a regime in which trajectories spend long times repeatedly returning to the well region, and a regime in which long jumps dominate the escape dynamics, effectively washing out the influence of the metastable basin.

Two limiting behaviors provide useful scaling intuition.
For large noise intensities, the MRT follows a power-law decay,
\begin{equation}
\tau_{\mathrm{MRT}}(x_0)\sim D_1^{-1},
\label{eq:largeD_review}
\end{equation}
consistent with escapes dominated by frequent large jumps and with the broader landscape of L\'evy-driven barrier crossing~\citep{Che05,Che07}.
In the opposite limit $D_1\to 0$, the MRT approaches finite, boundary-dependent values, reflecting the fact that rare but finite-probability long jumps can still induce barrier crossing even at vanishing noise intensity.

\medskip
Overall, the MRT framework provides a clear and robust characterization of metastable trapping under L\'evy flight dynamics. General equations enabling the exact calculation of the MRT for superdiffusive processes with arbitrary stability index $\alpha$ and arbitrary potential landscapes have been derived. In particular, the Cauchy case represents a rare setting in which non-Gaussian barrier crossing can be analyzed without uncontrolled approximations, and in which an NES enhancement naturally emerges in a form that is directly linked to experimental measurement protocols and boundary conditions.

\section{Switching dynamics in memristors}
\addcontentsline{toc}{section}{Memristor system}
\label{sec:memristors}

\begin{figure}[t]
\centering
\includegraphics[width=\columnwidth]{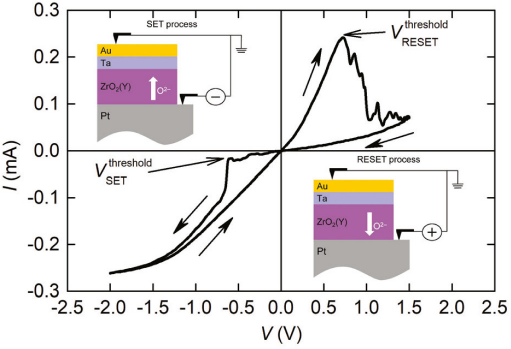}
\caption{Typical I--V curve of a memristive device based on a Ta/ZrO$_2$(Y)/Pt stack. The direction of the voltage sweep is shown by arrows. Inset: schematic representation of the oxygen ions drift under the action of an electric field in different directions. From \cite{Koryazhkina2022}.}
\label{fig:memristor1}
\end{figure}

Memristors are nonlinear, history-dependent two-terminal devices whose instantaneous resistance, memristance, depends on the time integral of the applied voltage or current, thereby providing an intrinsic mechanism for nonvolatile memory. Originally postulated by Chua as the fourth fundamental circuit element linking charge and magnetic flux~\citep{Chua1971}, the memristor completes the symmetry among the fundamental circuit variables. The modern physical realization of memristive behavior was experimentally demonstrated by \cite{Strukov2008} in TiO$_2$-based nanoscale devices, where resistance switching emerges from field-driven ionic transport and the formation of conductive filaments. From a dynamical-systems perspective, memristive systems are governed by coupled nonlinear equations with internal state variables, leading to a rich variety of phenomena, including hysteresis in the current--voltage characteristics, multistability, threshold dynamics, strong nonlinearity, and resistive switching (RS) between low-resistance (LRS) and high-resistance (HRS) states (see Fig.~\ref{fig:memristor1}).

These properties underpin the suitability of memristors for neuromorphic architectures, where they can emulate synaptic plasticity mechanisms. Moreover, their scalability to the nanoscale, low-energy switching, and compatibility with CMOS technologies make memristors promising candidates for high-density nonvolatile memories and in-memory computing platforms~\citep{Yang2013}.

Over the past two decades, resistive switching in memristive devices has attracted significant research attention~\citep{Shi2021,Sun2019}. Resistive switching (RS) refers to the bistable---or more generally multistable---transition of the resistance of a thin dielectric film sandwiched between two conductive electrodes under the application of an external voltage. To date, the prevailing understanding of the RS mechanism is largely based on the filamentary model, which attributes switching to the formation (set) and rupture (reset) of conductive filaments (CFs) connecting the electrodes within the functional dielectric layer under an applied electric field~\citep{Sun2019}.

\begin{figure}[t]
\centering
\includegraphics[width=0.85\columnwidth]{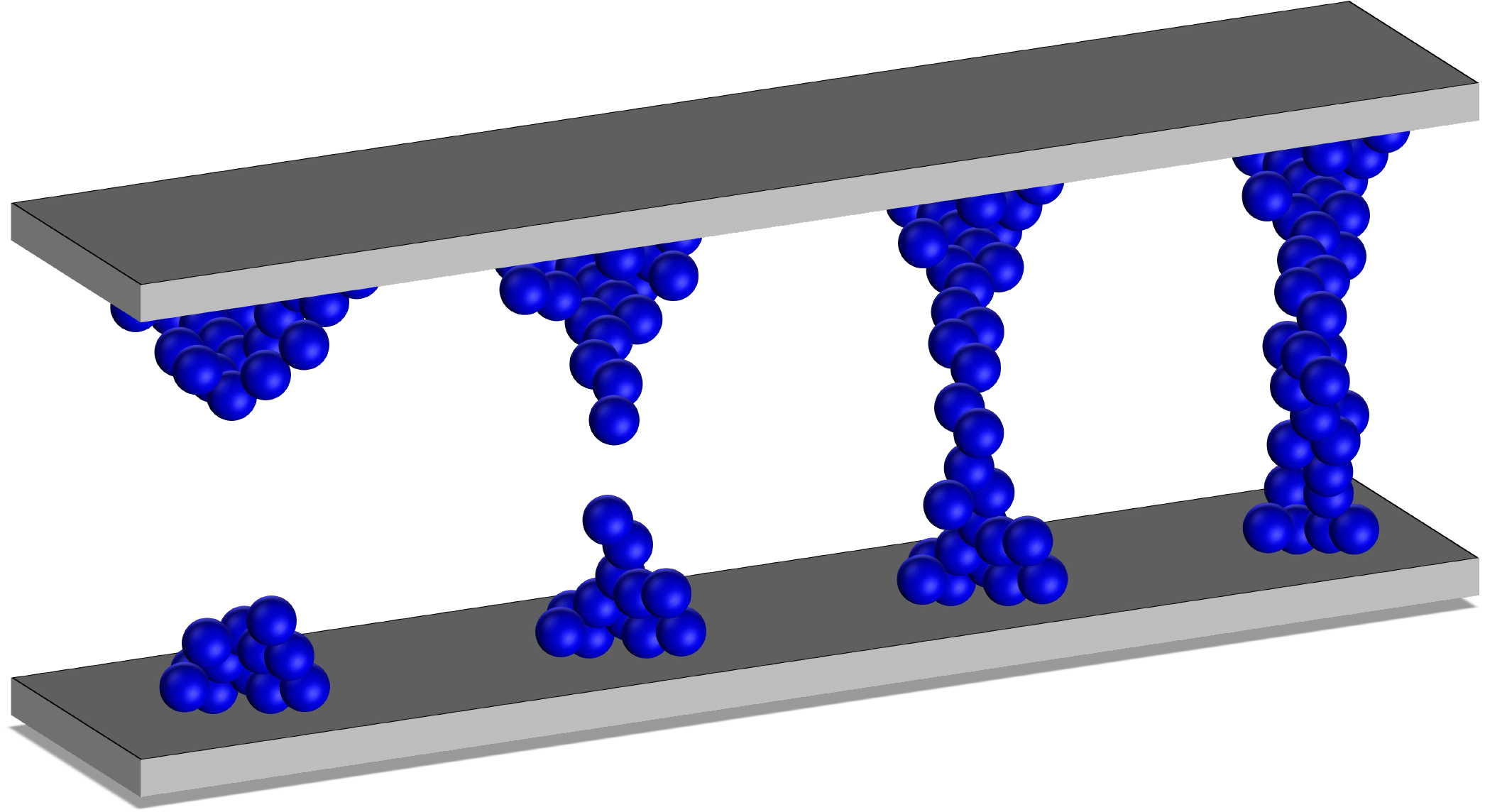}
\caption{A filament growth model for resistive random access memory switching.}
\label{fig:memristor2}
\end{figure}

In oxide-based memristive devices, conductive filaments are primarily composed of chains of oxygen vacancies. Memristors are therefore regarded as promising candidates for a wide range of applications, including nonvolatile resistive random access memory (RRAM), emerging computing architectures, and neuromorphic electronic circuits~\citep{Yu2016,Khan2021,Chua2022,Li2018,Boybat2018,Guo2022,Pi2019,Wang2021a,Wang2019,Wang2021b,Lv2018,Mahata2020,Wang2021c,Lv2020,Huang2021,Yu2013,Kim2015,Hussain2019,Mikhaylov2020,Resheed2021,Kousar2021,Alsuwian2021,Rasheed2021,Khera2022,DeStefano2024}. However, progress toward practical deployment is currently limited by the insufficient stability and reproducibility of resistive switching (RS) parameters during device operation~\citep{Alonso2021,Ielmini2016,Wu2014}.

Although substantial advances have been achieved in recent years, the overall performance of memristive devices remains inadequate for most large-scale applications. A key challenge is the intrinsically stochastic nature of the RS process, which constitutes a fundamental property of filament formation and rupture and leads to variability in switching voltages, resistance states, and device endurance~\citep{Koryazhkina2022}.

The stochastic nature of resistive switching has been widely observed and analyzed in the literature~\citep{Lee2015,Frick2022,Stotland2012,Gaba2013,Naous2016,Guarcello2017,Filatov2019,Ntinas2019,Naous2021,Ntinas2021,Wang2022,Patterson2013}. In this context, a novel strategy for improving memristor stability has recently been proposed. This approach treats the memristor as a multistable stochastic system and exploits well-established phenomena associated with the constructive role of noise in such systems. Within this framework, the device dynamics can be effectively described by a model of overdamped Brownian motion in a multistable force landscape, as proposed in \cite{Agudov2020}.

\begin{figure}[t]
\centering
\includegraphics[width=0.85\columnwidth]{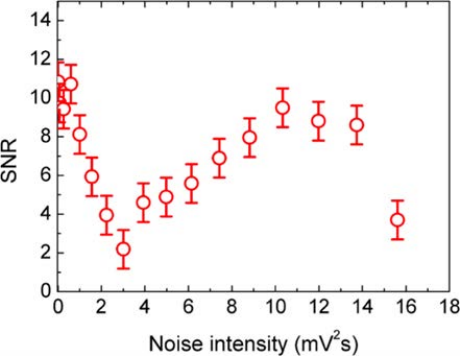}
\caption{Memristance SNR vs. noise intensity experimentally obtained for the memristive device under study. From~\cite{Mikhaylov2021}.}
\label{fig:memristor3}
\end{figure}

Several phenomena characteristic of multistable stochastic systems have been experimentally observed and theoretically investigated in memristive devices, including stochastic resonance~\citep{Mikhaylov2021}, resonant activation, and transient bimodality~\citep{Koryazhkina2022}. In particular, stochastic resonance has been studied both experimentally and theoretically in a metal-oxide memristive device based on yttria-stabilized zirconium dioxide and tantalum pentoxide, which exhibits bipolar filamentary resistive switching of anionic type. The effect of additive white Gaussian noise superimposed on a subthreshold sinusoidal driving signal was analyzed using time-series statistics of the resistive switching parameters, spectral response to periodic perturbations, and the output signal-to-noise ratio of the nonlinear system (see Fig.~\ref{fig:memristor3}).

These studies revealed stabilized resistive switching and an enhanced memristive response at an optimal noise intensity, consistent with the occurrence of stochastic resonance. The results were interpreted using a stochastic memristor model that explicitly accounts for an external noise source added to the control voltage. Overall, these findings clearly demonstrate that noise and fluctuations can play a constructive role in nonlinear memristive systems far from equilibrium~\citep{Mikhaylov2021}.

\begin{figure}[t]
\centering
\includegraphics[width=0.85\columnwidth]{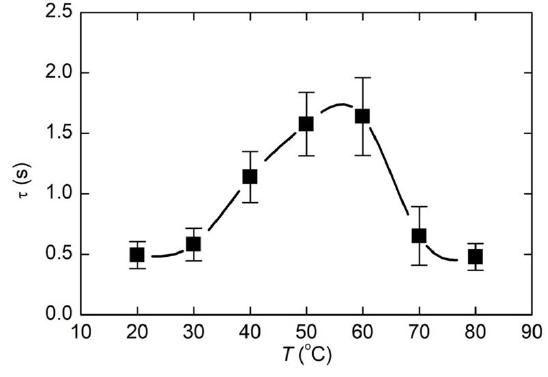}
\caption{Temperature dependence of the relaxation time of the memristor during the switching from LRS to HRS. From~\cite{Filatov2022}.}
\label{fig:memristor4}
\end{figure}

The phenomenon of noise-enhanced stability (NES) provides another striking example of the beneficial role of noise in stochastic nonlinear systems. The escape time $\tau$ of a Brownian particle in a fluctuating metastable state was investigated by \cite{Agudov2003} and \cite{Dub04}, where its dependence on the fluctuation intensity $D$ was shown to be nonmonotonic, exhibiting a maximum at an intermediate value of $D$. More recently, this effect has been observed experimentally in memristive devices~\citep{Filatov2022}. In that study, the kinetics of resistive switching (RS) during the transition from the low-resistance state (LRS) to the high-resistance state (HRS) was investigated as a function of temperature, which serves as a measure of the internal (thermal) noise intensity. The relaxation time $\tau$ displayed a nonmonotonic dependence on temperature, with a pronounced maximum. These experimental results were interpreted as evidence of the NES effect in memristors and constitute the first observation of noise-enhanced stability in memristive devices driven by thermal noise (see Fig.~\ref{fig:memristor4}).

Very recently, a nonmonotonic dependence of the relaxation time in a ZrO$_2$(Y)-based memristive device was observed during switching from the low-resistance state (LRS) to the high-resistance state (HRS) as a function of the intensity of an externally applied noise signal. The observed resistive switching kinetics exhibits at least two distinct regimes: a fast switching process from the LRS to an intermediate state with higher resistance, followed by a slower relaxation toward the stationary HRS. The occurrence of fast switching at random times is a hallmark of transient bimodality, a phenomenon that has been theoretically investigated in memristive systems using the model of overdamped Brownian motion in a multistable potential landscape~\citep{Agudov2020,Koryazhkina2022}. From a physical perspective, the fast switching regime corresponds to the initial rupture of the conductive filament.

\begin{figure}[t]
\centering
\includegraphics[width=0.85\columnwidth]{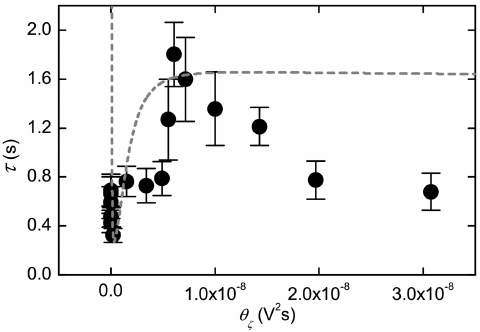}
\caption{Theoretical (dashed gray line) and experimental (black dots) dependencies of the relaxation time from LRS into HRS as a function of the external noise intensity $\theta_\zeta$. From~\cite{Koryazhkina2022}.}
\label{fig:memristor5}
\end{figure}

The dependence of the relaxation time $\tau$ on the external noise intensity $\theta_\zeta$ is shown in Fig.~\ref{fig:memristor5}. The behavior is clearly nonmonotonic, with a maximum occurring at $\theta_\zeta \approx 6.1\cdot 10^{-9}\,\mathrm{V^2 s}$. A simple stochastic model describing the dynamics of such a memristive device is based on a Langevin equation that accounts for the random hopping of a positively charged oxygen vacancy between trapping sites in the dielectric layer~\citep{Agudov2020}
\begin{equation}
\frac{dx}{dt} = - \frac{\partial \Phi(x)}{\partial x} + F + \xi(t),
\label{eq:mem_langevin}
\end{equation}
where $x$ is the coordinate of the hopping particle, $\xi(t)$ is a white Gaussian noise with statistical properties $\langle \xi(t)\rangle = 0$ and $\langle \xi(t)\xi(t+t')\rangle = 2\theta_\xi\,\delta(t')$, with $\theta_\xi = \eta k_B T$, where $k_B$ is Boltzmann constant, $\eta$ is viscosity coefficient according to the Sutherland--Einstein relation, and $T$ is the temperature. The potential landscape experienced by the hopping particle is modeled as a periodic potential $\Phi(x)$, consisting of potential wells separated by barriers of height $E_a$, corresponding to the activation energy. The particle dynamics is influenced by both the internal microstructure of the device and the externally applied electric field. The electric force acting on the particle is taken to be proportional to the applied voltage $F=qV/L$, where $q$ is the particle charge and $L$ is the effective distance between electrodes. If an additive noise source $\zeta(t)$ is applied to the driving voltage, the total voltage becomes
\begin{equation}
V = V_0 + \zeta(t),
\label{eq:mem_voltage}
\end{equation}
which leads to a stochastic modulation of the force term in Eq.~\eqref{eq:mem_langevin}, obtaining
\begin{eqnarray}\nonumber
\eta \frac{dx}{dt} =& - \frac{\partial \Phi(x)}{\partial x} + \frac{q}{L}V_0 + \frac{q}{L}\zeta(t) + \xi(t)\\
=& - \frac{\partial \Phi(x)}{\partial x} + \frac{q}{L}V_0 + \nu(t),
\label{eq:mem_effective}
\end{eqnarray}
where the total effective white Gaussian noise acting on the system is $\nu(t) = (q/L)\zeta(t) + \xi(t)$ with corresponding noise intensity $\theta_\nu = (q/L)^2\theta_\zeta + \theta_\xi$. In our experiment $\zeta(t)$ is a Gaussian noise with zero mean $\langle \zeta(t)\rangle = 0$ and noise intensity $\theta_\zeta = \sigma^2\tau_{\mathrm{corr}}$, where $\sigma^2$ is the variance and $\tau_{\mathrm{corr}}$ is the correlation time. For such a system, it has been theoretically demonstrated that the relaxation time can exhibit a nonmonotonic dependence on the noise intensity~\citep{Agudov2020}
\begin{equation}
\tau(\theta_\nu) = \left(\frac{L^2}{D_{\mathrm{eff}}}\right)\cdot \left[\frac{1}{\pi^2 + \gamma^2}\right],
\label{eq:mem_tau}
\end{equation}
which includes the effective diffusion coefficient $D_{\mathrm{eff}}$, the Kramers time $\tau_{\mathrm{kr}}=\tau_0\exp[E_a/\theta_\nu]$, and the dimensionless parameter $\gamma=(v_{\mathrm{eff}}L)/(2D_{\mathrm{eff}})$, with $v_{\mathrm{eff}}$ the effective drift coefficient. Explicit expressions for these quantities are provided in \cite{Koryazhkina2022}. We assume that the temperature remains constant throughout the experiment; consequently, the thermal noise intensity can be treated as constant. The theoretical dependence of the relaxation time $\tau$ (Eq.~\eqref{eq:mem_tau}) on the external noise intensity $\theta_\zeta$ is shown in Fig.~\ref{fig:memristor5} as a dashed gray curve~\cite{Agudov2020}, together with the experimental data (black dots). The agreement between theory and experiment is quite good, particularly with regard to the nonmonotonic behavior and the presence of a clear maximum in both curves.

This work provides the first experimental evidence of the noise-enhanced stability (NES) effect in the relaxation time dynamics of a memristive device induced by an external noise source. By contrast, the first experimental observation of NES in memristors driven by thermal (internal) noise was reported only very recently~\citep{Filatov2022}. These findings highlight the constructive role of both internal and external noise sources and point to their potential use as novel mechanisms for controlling resistive switching dynamics. In practical terms, externally applied noise offers a particularly attractive control strategy, as it can be more easily tuned and implemented to improve the stability and reproducibility of resistive switching.

\section{Noise enhanced phenomena in a quantum bistable system}
\addcontentsline{toc}{section}{NES in quantum bistable systems}

In this section, we explore the role of external driving in the escape dynamics of quantum metastable states under the influence of dissipation~\citep{Valenti2015,Valenti2018}. Understanding how a quantum system transitions out of a
metastable state is crucial for a range of applications, from
quantum information processing to condensed matter physics. In
particular, the interplay between monochromatic driving and
system--environment coupling can lead to complex behaviors. We find that, for appropriately chosen driving amplitudes, the escape time exhibits resonant peaks and dips as a function of the driving frequency when the dissipation is relatively weak. As the coupling strength increases, the escape time reaches a pronounced maximum, followed by a rapid decline at a critical value of the coupling. This behavior marks a crossover to a regime in which the escape time becomes essentially independent of the driving frequency, closely resembling the behavior observed in the static (undriven) case. These results highlight the rich dynamics arising from the combined effects of driving and dissipation in quantum metastable systems.

\subsection{Caldeira-Leggett model}

A paradigmatic example of a quantum dissipative system is given by the Caldeira--Leggett model~\citep{CaldeiraLeggett1983}, in which a system $S$ with
generalized coordinate $\vec{x}$ is coupled to a dissipative
environment, represented by a heat bath of quantum harmonic
oscillators with frequencies $\omega_j$ and coordinates $\vec{x}_j$ through the Hamiltonian 
\begin{equation}
\hat{H}(t)\!=\!\hat{H}_{S}(t)\!+\!\frac{1}{2}\!\sum_{j=1}^{N}\!\!\left[\frac{\hat{p}_{j}^{2}}{m_{j}}\!+\!m_{j}\omega_{j}^{2}\!\left(\!\hat{x}_{j}\!-\!\frac{c_{j}\hat{x}}{m_{j}\omega_{j}^{2}}\right)^{\!\!2}\right]\label{eq01PRA}
\end{equation}
with $H_S(t)$ accounting for the presence of a time-varying potential. The second term on the right-hand side of 
Eq.~(\eqref{eq01PRA}) contains the free bath Hamiltonian, the bilinear interaction term between S and the oscillators, and a renormalization term, dependent on $\vec{x}$. This term ensures a purely dissipative bath, resulting in dissipation that is independent of the coordinate of the central system. The resulting time-dependent Hamiltonian for S reads
\begin{equation}
\begin{aligned}
\hat{H}_{S}(t) &= \frac{\hat{p}^{2}}{2M} + V(\hat{x}) - \hat{x}\,A\sin(\Omega t) \\
               &= \hat{H}_{0} - \hat{x}\,A\sin(\Omega t).
\end{aligned}
\end{equation}
\enspace where the static potential $\vec{V}(x)$ is given by
\begin{equation}
V(\hat{x})=\frac{M^{2}\omega_{0}^{4}}{64\,\Delta U}\,\hat{x}^{4}-\frac{M\omega_{0}^{2}}{4}\,\hat{x}^{2}-\epsilon\,\hat{x}.
\end{equation}
When the bath is assumed to be in thermal equilibrium, its effect on the quantum bistable system is completely
described by the Ohmic spectral density~\cite{Weiss2012,Magazzu2015}.

\subsection{Discrete variable representation}

\begin{figure}
    \centering
    \includegraphics[width=0.85\linewidth]{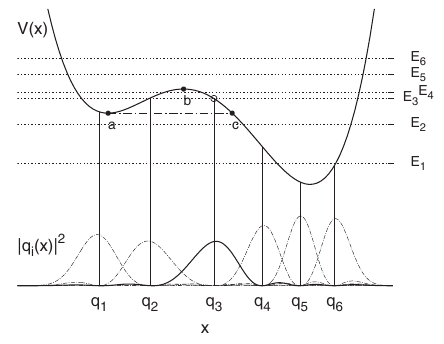}
\caption{Potential $V$ [Eq.~(3), with $\Delta U = 1.4\hbar\omega_{0}$ and $\epsilon = 0.27\sqrt{M\hbar\omega_{0}^{3}}$] and the first six energy levels (horizontal lines). In the lower part, the probability densities $|q_i(x)|^{2}=|\langle x|q_i\rangle|^{2}$ associated with the DVR eigenfunctions are shown, the initial state $|q_{3}\rangle$ being highlighted by a solid line. Vertical lines indicate the position eigenvalues in the DVR. The metastable region of the potential is to the left of the so-called \emph{exit point} $c$. From~\cite{Valenti2018}.}
    \label{fig:potential}
\end{figure}

At low temperatures, relative to the energy scale set by $\omega_0$, the particle's time evolution is effectively restricted to a reduced Hilbert space spanned by the first M energy eigenstates $|E_i>$, provided that the particle is not initially excited to levels above M. We further assume that the time-dependent driving does not induce transitions to energy eigenstates beyond those already considered in
the static potential. In addition, these energy eigenstates are the same as those of the system described by $\vec{H}_0$ in Eq.~(2). Moreover, when the frequency of the periodic external driving is comparable to or larger than the energy gap between the ground state and the first excited state, $\Omega \geq \omega_0$, it is appropriate to average the dynamics over a full driving period~\citep{Grifoni1998,Thorwart2000,Thorwart2001}. After this averaging, the resulting transition coefficients
become time-independent, forming the time-averaged rate matrix. If the driving frequency does not satisfy this high-frequency condition, Floquet theory provides an alternative theoretical
framework.

The populations $\rho_{jj}$ of the discrete variable representation~\citep{Feynman1963,Harris1965,LightCarrington2007} (DVR) states $|q_j>$ relax toward a stationary configuration that depends on the bath parameters and the damping strength $\gamma$. Consequently, at strong coupling, this
relaxation process is well described by the incoherent dynamics
governed by the master equation~\citep{Thorwart2000,Thorwart2001,Magazzu2015}
\begin{equation}\label{Eq15PRA}
\dot{\rho}_{jj}(t)=\sum_{k}\Gamma^{\mathrm{av}}_{jk}\,\rho_{kk}(t),
\end{equation}
with $\Gamma^{av}_{jk}$ being time-independent averaged rates.

\subsection{Escape time for the driven system}

We study the transient dynamics of the driven system described by Eq.~\eqref{Eq15PRA}, starting from a nonequilibrium initial condition,
\begin{equation}
    \rho(0)=|q_3><q_3|,
    \label{Eq22PRA}
\end{equation}
\begin{figure}
    \centering
    \includegraphics[width=\linewidth]{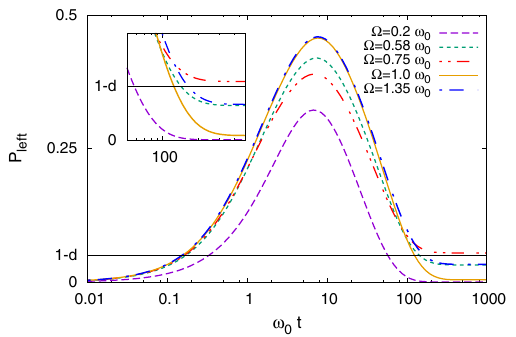}
    \caption{Left well population $P_{left} = \rho_{11} + \rho_{22}$ as a function of time (in log scale) for $\bar{A} = 0.15, \bar{T} = 0.1$, $\gamma/\omega_0$ = 0.2, and five driving frequency values, namely $\Omega/\omega_0$ = 0.2,0.58,0.75,1.0,1.35. Here we introduced the
    dimensionless quantities of the amplitude and temperature, $\bar{A}=A/\sqrt{M \hbar \omega_0^3}$ and
    $\bar{T}=k_B T/\hbar \omega_0$. Horizontal line: 1-d, where d = 0.95 is the threshold value. For $\Omega/\omega_0$ = 0.75 no escape occurs, as $P_{left}$ remains above the value 1-d.
    This is better shown in the inset. Note the nonmonotonic behavior of the steady-state values of $P_{left}$ as a function of $\Omega$. From~\cite{Valenti2018}.}
    \label{fig:resEXP-BGC1}
\end{figure}
that is, with the particle initially localized in the central region of the potential on the right side of the barrier, between the maximum and the exit point, denoted as c in Fig.~\ref{fig:potential}. Consequently, the time evolution of the populations in our asymmetric bistable quantum system (see Fig.~\ref{fig:resEXP-BGC1}), under the initial condition Eq.~\eqref{Eq22PRA}, is described by 
\begin{equation}
\rho_{ij}(t)=\sum_{n=1}^{6} S_{jn}\,(S^{-1})_{n3}\,e^{\Lambda_{n}(t-t_{0})}\,\rho_{33}(t_{0}) .
\end{equation}
We analyze the escape time from the metastable region, defined as the area to the left of the exit point c in Fig.~\ref{fig:potential}, following ~\cite{Sargsyan2007}, where the decay rate from the metastable region is determined using the probability of a Gaussian wave packet tunneling from left to right across the potential barrier shown in Fig.~\ref{fig:potential}. In the present work,
we employ a discretized version of this method. Specifically, we compute the population of the lower (right-side) well, corresponding to the cumulative population of the three DVR states $|q_4>$,
$|q_5>$, and $|q_6>$,
\begin{equation}
P_{\mathrm{right}}(t)=\sum_{j=4}^{6}\rho_{jj}(t).
\end{equation}
\enspace We define the escape time $\tau$ from the metastable region of the potential (the area to the left of the exit point c) as the time required for the population of the right well to reach a threshold value d. In the static case, \cite{Valenti2015} showed that the nonmonotonic behavior of $\tau$ as a function of $\gamma$ and the temperature $T$ remains robust under small variations of the threshold
around $0.9$. In the present study, we set the threshold to $d=0.95$, meaning that the particle is considered to have escaped from the metastable region once the probability of finding it in the lower (right) well reaches or exceeds 95\%.

It is worth noting that, in the static case, the metastable well can be thermally populated at the steady state. In such a scenario, no escape occurs if the threshold d is close to unity~\citep{Valenti2015}. As we will show below, a similar situation arises in the driven case for certain driving frequencies $\Omega$, particularly at large amplitudes A, when the left-well population, given by
$P_{left}=\rho_{11}+\rho_{22}$, remains significantly above zero at the steady state due to the effect of the driving.

\begin{figure}
    \centering
    \includegraphics[width=\linewidth]{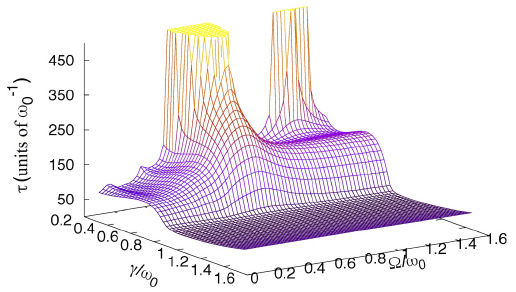}
    \caption{Escape time as a function of the coupling strength and the driving frequency for dimensionless amplitudes $\bar{A} = 0.15$.
    The dimensionless temperature is set to the value $\bar{T} = 0.1$. From~\cite{Valenti2018}.}
    \label{fig:resEXP-BGC2}
\end{figure}

\begin{figure}
    \centering
    \includegraphics[width=\linewidth]{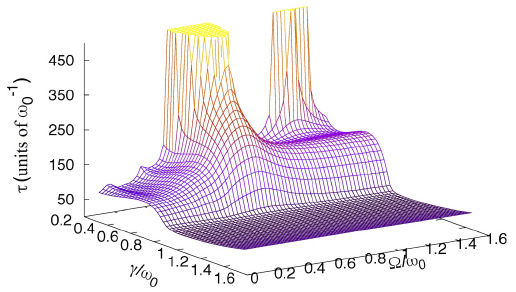}
    \caption{Escape time as a function of the coupling strength and the driving frequency for dimensionless amplitudes $\bar{A} = 0.20$.
    The dimensionless temperature is set to the value $\bar{T} = 0.1$. From~\cite{Valenti2018}.}
    \label{fig:resEXP-BGC3}
\end{figure}

An example of this effect is shown in Fig.~\ref{fig:resEXP-BGC1}, where, for $\bar{A}=0.15, \bar{T}=0.1$, and $\gamma/\omega_0$ = 0.2, the
population of the left well is plotted as a function of the time for
various driving frequencies, where the dimensionless amplitude
$\bar{A}=A/\sqrt{M \hbar \omega_0^3}$ and temperature $\bar{T}=k_B T/\hbar \omega_0$. The stationary value of $P_{left}$ remains above
$1-d=0.05$ for $\Omega/\omega_0$ = 0.75, indicating that no escape
occurs at this driving frequency. 
\begin{equation}
\rho_{jj}(t)=\sum_{n,k=1}^{M} S_{jn}\,(S^{-1})_{nk}\,e^{\Lambda_{n}(t-t_{0})}\,\rho_{kk}(t_{0}),
\end{equation}
Our primary objective is to study how the escape time depends on the driving frequency and the coupling strength. To this end, we first fix $\gamma$ at the value used in Fig.~\ref{fig:resEXP-BGC1}, which is the lowest value
considered in this work, and plot the escape time as a function of both $\Omega/\omega_0$ and $\gamma/\omega_0$ for two different amplitude values, $\bar{A} =
0.15$ and $0.20$, respectively, see Figs.~\ref{fig:resEXP-BGC2} and~\ref{fig:resEXP-BGC3}.
As $\gamma$ increases, the peaks and dips of $\tau$ as a function of the driving frequency gradually smooth out, until $\tau$ becomes nearly independent of $\gamma$ as the coupling approaches a critical value $\gamma_c = 0.75$. For $\gamma<\gamma_c$, high-frequency driving can hinder or enhance the escape process, while for coupling strengths exceeding the critical value $\gamma_c$, the escape time no longer depends on the driving frequency and the quantum Zeno effect sets in \cite{Facchi2005}.

The critical value $\gamma_c$ defines the onset of a dynamical
regime in which the tunneling process responsible for the
transfer of the population to the states of the metastable region is suppressed.
This occurs because the tunneling dynamics slows down relative to the depletion time scale of the region where the particle is initially localized. Consequently, the probability of finding the particle in the metastable well, starting from the initial condition~\ref{Eq22PRA}, is effectively zero at all times, as the population is transferred directly from $|q_3>$ to the states of the right well.
%

\section{Axion-Induced Resonant Activation in Josephson Switching Dynamics}
\addcontentsline{toc}{section}{JJ-based axion detection: brief review note}

We outline a switching-statistics strategy for axion detection using a current-biased Josephson junction (CBJJ) operated as a threshold device. If an axion field couples to the Josephson phase, it can act as an additional weak effective drive and modify the escape dynamics from the superconducting to the finite-voltage state. The central predicted signature is an axion-induced resonant-activation effect, observed as a pronounced minimum in the mean switching time when the tunable junction energy scale is varied. Moreover, a similar structure in the width of the switching-time distributions appears.

\subsection{Motivation and statistical viewpoint} \label{sec:intro}

Josephson junctions, and more broadly superconducting weak-link devices, are widely used as sensitive detectors and readout elements~\citep{Bar82,Dev13,Taf19,Bra19,Krantz,Citro2024}.
They combine strong nonlinearity with low dissipation and low intrinsic noise, and they offer a direct electrical readout in modern superconducting-circuit platforms~\citep{Dev13,Kja20}.
In many implementations a JJ acts as a threshold element, where small perturbations (current, magnetic flux, microwave drives, or other weak effective forces on the phase) can produce a pronounced change in the phase dynamics, up to triggering the switch from a metastable superconducting state to a finite-voltage state \citep{Bar82,Dev84,Guarcello16PRB}.
This ``switching-based'' strategy is attractive because it can convert weak inputs into measurable large voltage pulses, and it enables statistically robust inference through repeated trials \citep{Dev84,Blackburn2016,Guarcello21APL,Guarcello2023APL,Guarcello2024APL}.
Similar ideas underlie a broader landscape of superconducting detectors, including devices operating in regimes relevant to photon and bolometric detection, where low noise and strong nonlinearity are exploited to transduce extremely weak excitations into measurable electrical signals \citep{Chen2011PRL,Oelsner17PRAppl,Gua19,Lee20,Revin2020,Wal21,Ret21,Pankratov2025}.
Moreover, Josephson devices have also been discussed in broader contexts at the interface between superconductivity and axion or axion-like physics, both from the standpoint of particle searches and of axion-electrodynamics-induced Josephson phenomena \citep{Asztalos2010PRL,Ira18,Nog16,Kuzmin2018IEEETAS,Pankratov2022,Sushkov2023,Bartram2023,Braggio2025PRX}.

Within this detector paradigm, a recent and debated line of work has suggested that switching dynamics in a CBJJ could also be sensitive to axion-like dark matter~\citep{Bec11,Bec12,Bec13,Bec15,Bec16}.
The basic hypothesis is that, under specific conditions, a background axion field may \emph{couple} to the Josephson phase, thus effectively behaving as an additional time-dependent drive acting on the junction.
If such a coupling is present, it should manifest as systematic, reproducible changes in the statistics of escape (switching) events from the superconducting to the resistive state.
This provides a basis for a switching-statistics viewpoint as a proposed protocol, in which one would test the hypothesis, i.e., the presence of the axion filed, by measuring distributions and their dependence on control parameters, such as mean values and variances, over a large ensemble of switching realizations~\citep{Gua19,Pie21,Gua21}.

The proposed framework focuses on switching-time statistics, namely, the time it takes for a junction biased below its critical current to leave the metastable superconducting state and reach the finite-voltage state~\citep{Grimaudo2022}. Under the coupling hypothesis, an axion-induced \emph{resonant activation} (RA) effect may appear as a minimum of the mean switching time (MST), i.e., the average switching time over many independent repeated realizations, as a function of a tunable energy-ratio parameter.
This research line is motivated by two considerations.
First, the axion misalignment angle obeys an equation of motion that is formally analogous to that of a damped pendulum, and thus resembles the nonlinear equation governing the Josephson phase dynamics in a biased junction~\citep{Sik83,Co20}.
Second, the ratio between the axion and the Josephson energy scales can, in principle, be tuned experimentally. A convenient control parameter is the dimensionless quantity $\varepsilon=(E_a/\hbar\omega_p)^2$, where
$E_a=m_a c^2$ (with $m_a$ the axion mass) and $\omega_p$ is the Josephson plasma frequency.
Since the latter depends on junction parameters and in particular on the critical current~\citep{Bar82}, one can scan $\varepsilon$ by varying the device working conditions, for instance through temperature, magnetic field, or electrostatic gating. This can be done either within a single junction or across an array of junctions with different critical currents~\citep{Dub01,Ber08,Du08,DeS19}.

Complementary studies have extended this research framework in two directions.
On the one hand, the influence of the axion field has been recast in terms of an effective (quasi)potential barrier governing escape, providing an alternative route to connect switching statistics to an inferred coupling strength~\citep{GrimaudoMaterials2023}.
On the other hand, in a low-noise quantum limit the coupled dynamics has been discussed in terms of interacting two-level systems, where the axion-like subsystem is indirectly revealed through induced oscillations of an accessible Josephson degree of freedom~\citep{GrimaudoCSF2023}.

In this perspective, if the coupling mechanism assumed in earlier works were realized in a given experimental configuration, one would expect distinctive statistical signatures to emerge in switching-time measurements. In particular, since the term $\propto \varepsilon\sin\theta$ acts as an effective ac-like drive in the JJ dynamics, $\langle\tau\rangle(\varepsilon)$ may exhibit RA and develop a minimum when a characteristic axion--JJ oscillation frequency matches the effective plasma frequency of the CBJJ~\citep{Grimaudo2022}.

Importantly, one can extend the analysis beyond the mean switching time by tracking the width of the switching-time distribution as well. Similar structures in both mean and dispersion through controlled $\varepsilon$ scans would provide a more solid and quantitative basis for supporting the axion-coupling hypothesis and for possible experimental validation of the phenomenon.

\subsection{Josephson phase dynamics and switching statistics}

Within the resistively and capacitively shunted junction (RCSJ) description, the Josephson phase $\varphi$ evolves as a damped particle in a tilted washboard potential~\citep{Stewart1968,McCumber1968,Bar82}.
In normalized units, i.e., $\tau_c=\omega_c t$, with $\omega_c=(2e/\hbar)I_cR$, the stochastic equation for a CBJJ reads
\begin{equation}
\beta_c\,\ddot{\varphi}(\tau_c)+\dot{\varphi}(\tau_c)+\sin\varphi(\tau_c)= i_b + i_n(\tau_c),
\label{eq:rcsj}
\end{equation}
where overdots denote derivatives with respect to $\tau_c$, $i_b=I_b/I_c$ is the dc bias current normalized to the critical current $I_c$, and $\beta_c=\omega_c R C$ is the Stewart--McCumber parameter, with $R$ and $C$ being the normal-state resistance and
capacitance of the JJ, respectively.
Thermal fluctuations are modeled as a Gaussian white current noise with $\langle i_n\rangle=0$ and
\begin{equation}
\langle i_n(\tau_c)\,i_n(\tau_c+\tilde{\tau}_c)\rangle = 2\Gamma\,\delta(\tilde{\tau}_c),
\label{eq:noise}
\end{equation}
where $\Gamma=\frac{k_B T}{R}\,\frac{\omega_c}{I_c^{2}}$ is the dimensionless noise intensity~\citep{Bar82}, with $k_B$ being the Boltzmann constant and $T$ the temperature.
For $i_b<1$ the washboard potential hosts metastable minima. Switching corresponds to noise-activated escape over the nearest barrier.
The key observables for our purposes are the average value, $\langle\tau\rangle$, and the width, i.e., the standard deviation, $\sigma_\tau$, of the switching-time distributions.\\

\subsection{Axion equation of motion and coupling to a JJ}

The axion misalignment angle $\theta$ obeys, in a homogeneous approximation, to the equation
\begin{equation}
\ddot{\theta}(t) + 3H\,\dot{\theta}(t) + \frac{m_a^2c^4}{\hbar^2}\,\sin\theta(t)=0,
\label{eq:axion_eom}
\end{equation}
where $m_a$ is the axion mass and $H$ is the Hubble parameter~\citep{Sik83,Co20}. 

\begin{figure*}[t]
\centering
\includegraphics[width=\textwidth]{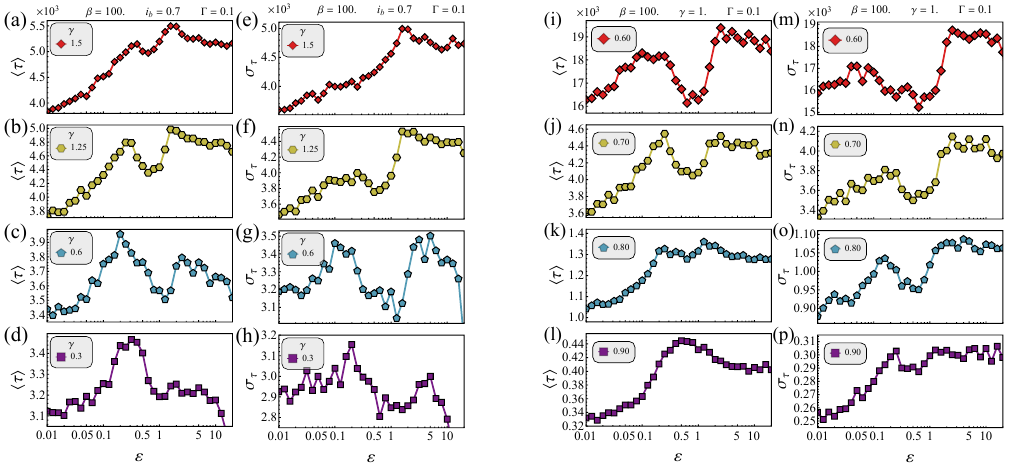}
\caption{Switching-time statistics as a function of the energy-ratio parameter $\varepsilon$, for an underdamped junction ($\beta=100$) at fixed noise intensity $\Gamma=0.1$.
Left columns: mean switching time $\langle\tau\rangle$, panels (a)–(d), and standard deviation $\sigma_\tau$, panels (e)–(h), versus $\varepsilon$ at fixed coupling $i_b=0.7$, for four axion--JJ coupling values, $\gamma=1.5$ (top row), $1.25$, $0.6$, and $0.3$ (bottom row).
Right columns: mean switching time $\langle\tau\rangle$, panels (i)–(l), and standard deviation $\sigma_\tau$, panels (m)–(p), versus $\varepsilon$ at fixed coupling $\gamma=1$, for four bias values, $i_b=0.60$ (top row), $0.70$, $0.80$, and $0.90$ (bottom row).}
\label{fig:AxionJJ_switching_stats}
\label{fig:4panel}
\end{figure*}

It is convenient to write the coupled axion--JJ dynamics in normalized units~\citep{Grimaudo2022}.
Using the characteristic-frequency normalization $\tau_c=\omega_c t$, with $\omega_c=(2e/\hbar)I_cR$, one obtains the system of stochastic equations
\begin{subequations}
\label{eq:B5_like}
\begin{align}
\beta_c\ddot{\varphi}
+ k_2\dot{\varphi}
+ k_2\sin\varphi
+ k_1\varepsilon\,\sin\theta
&= k_2\,(i_b+i_n),
\label{eq:B5_like_phi}\\
\beta_c\ddot{\theta}
+ k_1\dot{\varphi}
+ k_1\sin\varphi
+ k_2\varepsilon\,\sin\theta
&= k_1\,(i_b+i_n),
\label{eq:B5_like_theta}
\end{align}
\end{subequations}
where
\begin{equation}
k_1=\frac{\gamma}{1+2\gamma},
\qquad
k_2=\frac{1+\gamma}{1+2\gamma},
\end{equation}
with $\gamma$ being the coupling constant between axion and JJ.
An axion friction term proportional to $\dot{\theta}$ is neglected in this model, due to the extremely small ratio $3H/\omega_p$.
In addition to a dc bias and thermal noise in the JJ branch, the coupling introduces an additional effective term, $k_1\varepsilon\sin\theta$.
The central control parameter is the dimensionless \emph{energy ratio}
\begin{equation}
\varepsilon = \left(\frac{m_ac^2}{\hbar\omega_p}\right)^{\!2},
\qquad\text{with}\qquad
\omega_p=\sqrt{\frac{2eI_c}{\hbar C}}.
\label{eq:epsilon}
\end{equation}
A crucial experimental point is that the Josephson plasma frequency is experimentally tunable, because $I_c$ can be modified in situ by changing the device working point, for instance via temperature, applied magnetic field, or electrostatic gating~\citep{Dub01,Ber08,Du08,DeS19}.
This makes it possible to scan the energy-scale ratio $\varepsilon$ in a controlled manner.
Operationally, one can scan $\varepsilon$ either by tuning $I_c$ in situ in a given device, while keeping $i_b=I_b/I_c$ in a chosen range, or by using an array of junctions with different $I_c$ that directly sample different $\varepsilon$ values.

The system of Eqs.~\eqref{eq:B5_like} is integrated numerically using an explicit finite-difference scheme with time step $\Delta t=10^{-2}$.
Switching-time statistics are then computed over an ensemble of $10^{4}$ independent realizations.

\subsection{Results}

Figure~\ref{fig:4panel} shows that, in the underdamped regime ($\beta=100$) at fixed noise intensity $\Gamma=0.1$, the switching dynamics displays an axion-induced RA effect:
the mean switching time becomes a nonmonotonic function of $\varepsilon$ and exhibits a clear minimum in the interval $\varepsilon\in[0.1,1]$ for intermediate coupling values.
Physically, this minimum arises from a frequency-matching condition between the effective Josephson plasma oscillations and the characteristic frequency of the coupled axion--JJ linearized dynamics.
The term proportional to $\varepsilon\sin\theta$ can be interpreted as an oscillating drive acting on the Josephson phase~\citep{Grimaudo2022}, with characteristic frequencies obtained from the small-oscillation linearization of the coupled system (noise-free limit).
As in standard resonant-activation scenarios, the mean escape time becomes a nonmonotonic function of the control time scale associated with a periodic drive or a time-dependent barrier.
When this external time scale approaches the intrinsic oscillation time scales of the metastable dynamics, the escape process is favored, yielding a pronounced minimum in $\langle\tau\rangle$ as a function of the tuning parameter~\citep{Dev84,Doe92,Kau96,Gua15,Ladeynov2023CSF}.
In the same regime, one often expects a similar structure in the width of the switching-time distribution, which can be quantified by its standard deviation $\sigma$.

Panels (a–d) and (e–h) report, respectively, the mean switching time $\langle\tau\rangle$ and the standard deviation $\sigma_\tau$ versus $\varepsilon$ for several coupling strengths $\gamma$, from $0.3$ to $1.5$, at fixed bias $i_b=0.7$.
Changing $\gamma$ modifies both the switching times and the width of the distribution, and it also affect the visibility of the nonmonotonic structures in $\varepsilon$.

The right-hand set of panels focus on the role of the bias current at fixed coupling, $\gamma=1$.
In fact, panels (i)–(l) show $\langle\tau\rangle(\varepsilon)$ for four representative values of $i_b$, from $0.6$ to $0.9$, while panels (m)–(p) show the corresponding $\sigma_\tau(\varepsilon)$.
A clear and systematic trend is that increasing $i_b$ reduces the switching times, as expected for escape from a tilted metastable potential, while preserving a pronounced structure versus $\varepsilon$.
The nonmonotonic features are most visible at lower bias, where the metastable well is deeper and the escape time scale is longer, whereas at higher bias the curves become flatter and the relative modulation is reduced.
In the same $\varepsilon$ region where $\langle\tau\rangle$ displays a dip, $\sigma_\tau$ typically shows also a nonmonotonic modulation in the same $\varepsilon$ window, which is consistent with a reshaping of the full switching-time distribution rather than a simple rigid shift of its mean.

Overall, Fig.~\ref{fig:AxionJJ_switching_stats} provides a compact view of how the proposed signature, i.e., a reduction of the mean switching time due to RA, accompanied by a corresponding feature in the dispersion, depends on both the coupling parameter $\gamma$ and the operating point set by $i_b$.

Two additional qualitative features are emphasized as practical detection tools.
First, for $\varepsilon\ll 1$ the axion term becomes ineffective and Eqs.~\eqref{eq:B5_like} effectively decouple, so switching statistics approach the unperturbed JJ behavior.
Second, for $\varepsilon\gg 1$ the axion-induced term produces for sufficiently large $\gamma$ a plateau in the switching-time observables
Therefore, an experimental scan in $\varepsilon$ can simultaneously test for (\emph{i}) the presence of an RA minimum near $\varepsilon\lesssim 1$ and (\emph{ii}) a statistically significant separation between the small-$\varepsilon$ and large-$\varepsilon$ plateaus.
Finally, as discussed by \cite{Grimaudo2022}, we stress that the coupling-induced signatures are expected to be most visible in the underdamped regime; for strong damping (small $\beta_c$) the effective coupling becomes negligible, making this mechanism ineffective.

\section{Conclusions}

In this work we have reviewed a collection of recent results that share a common message: metastability in nonlinear systems can be controlled, and in specific regimes even reinforced, by fluctuations, dissipation, and time-dependent perturbations. Starting from L\'evy-driven barrier crossing, we emphasized how heavy-tailed noise qualitatively alters escape physics, yielding a finite, boundary-dependent residence-time behavior in the small-noise limit and producing a pronounced nonmonotonic dependence of the mean residence time on the L\'evy-noise intensity. This provides a rigorous and measurement-oriented extension of noise-enhanced stability beyond the Gaussian setting.

We then connected these ideas to solid-state devices, focusing on memristors as paradigmatic multistable nonequilibrium systems. In oxide-based filamentary devices, the intrinsically stochastic nature of resistive switching leads to large variability, but also enables constructive noise-induced phenomena. The experimental evidence of stochastic resonance and of noise-enhanced stability, both under internal (thermal) noise and under externally injected noise, supports the view that controlled fluctuations can become a practical knob for stabilizing operation, improving reproducibility, and shaping switching kinetics in neuromorphic and memory-oriented platforms.

A complementary quantum perspective emerges in driven dissipative bistable systems described within the Caldeira–Leggett framework. There, the escape dynamics from a metastable region exhibits a crossover controlled by system–bath coupling: from a regime where driving frequency produces resonant peaks and dips in the escape time, to a strongly dissipative regime where the escape becomes essentially frequency independent and the dynamics is dominated by dissipation-controlled depletion pathways. This behavior constitutes a quantum analogue of noise-assisted control, showing that dissipation, when properly engineered, can be used to tailor metastable lifetimes rather than merely destroying coherence.

Finally, we discussed an application at the interface of nonlinear dynamics and fundamental physics: axion-induced modifications of Josephson switching statistics. Within the effective coupling hypothesis, scanning the ratio between axion and Josephson energy scales can generate a minimum in the mean switching time, accompanied by a similar modulation of the distribution width. Independently of the broader debate on microscopic coupling mechanisms, this framing suggests a concrete statistical protocol: repeated switching measurements, combined with controlled scans of junction parameters, can be used to search for reproducible, parameter-locked signatures that go beyond a shift of the mean and involve the full switching-time distribution. 

Overall, the examples reviewed here show that metastability provides a unifying language across classical, quantum, and device-level settings, and that noise-assisted phenomena can be leveraged both to interpret complex dynamics and to design robust control and sensing strategies.



\begin{dci}
The author(s) declared no potential conflicts of interest with respect to the research, authorship, and/or publication of this article.
\end{dci}

\begin{funding}
This work was partially supported by Italian Ministry of University and Research (MUR) and National Center for Physics and Mathematics (section No. 9 of scientific program “Artificial intelligence and big data in technical, industrial, natural and social systems”). D.V. acknowledges support from European Union - Next Generation EU through project THENCE - Partenariato Esteso NQSTI - PE00000023 - Spoke 2.
\end{funding}



\end{document}